\documentclass[aps,prl,showpacs,twocolumn,superscriptaddress,floatfix]{revtex4}

\usepackage{graphicx}
\usepackage{amsmath}
\usepackage{amssymb}
\usepackage{bm}
\usepackage{dcolumn}

\begin{document}
\title{Electronic structure induced reconstruction and magnetic
ordering at the LaAlO$_3|$SrTiO$_3$ interface}
\author{Zhicheng Zhong}
\affiliation{Faculty of Science and Technology and MESA$^+$
Institute for Nanotechnology, University of Twente, P.O. Box 217,
7500 AE Enschede, The Netherlands}
\author{Paul J. Kelly}
\affiliation{Faculty of Science and Technology and MESA$^+$
Institute for Nanotechnology, University of Twente, P.O. Box 217,
7500 AE Enschede, The Netherlands}
\date{\today}
\begin{abstract}
Using local density approximation (LDA) calculations we predict
GdFeO$_3$-like rotation of TiO$_6$ octahedra at the $n$-type interface
between LaAlO$_3$ and SrTiO$_3$. The narrowing of the Ti $d$ bandwidth
which results means that for very modest values of $U$, LDA$+U$
calculations predict charge and spin ordering at the interface. Recent
experimental evidence for magnetic interface ordering may be understood
in terms of the close proximity of an antiferromagnetic insulating
ground state to a ferromagnetic metallic excited state.
\end{abstract}

%
% 68.35.-p Solid surfaces and solid�solid interfaces: structure and
%          energetics
%
% 68.35.Ct Interface structure and roughness
%
% 71.28.+d Narrow-band systems; intermediate-valence solids
%          (for magnetic aspects, see 75.20.Hr and 75.30.Mb
%           in magnetic properties and materials)
%
% 73.20.-r Electron states at surfaces and interfaces
%
% 73.20.Hb Impurity and defect levels; energy states of adsorbed species
%
% 73.40.-c Electronic transport in interface structures
%
% 75.70.Cn Magnetic properties of interfaces (multilayers,
%          superlattices, heterostructures)
%

\maketitle

Transition metal (TM) oxides in bulk form exhibit a huge range of
physical properties. Heterostructures of TM oxides offer the prospect
of greatly enhancing these properties or of combining them to realize
entirely new properties and functionalities. The recent finding that a
TiO$_2|$LaO interface between the insulating oxides LaAlO$_3$ and
SrTiO$_3$ can be metallic with an extremely high carrier mobility
\cite{Ohtomo:nat04} has triggered a surge of experimental and
theoretical studies of this interface
\cite{Huijben:natm06,Thiel:sc06,Brinkman:natm07,Reyren:sc07,
Herranz:prl07,Siemons:prl07,Willmott:prl07,Kalabukhov:prb07,
Basletic:cm07,Pentcheva:prb06,Park:prb06}. The valence mismatch at the
interface leads to the transfer of half an electron per unit cell from
LaAlO$_3$ (LAO: bandgap 5.6~eV) to SrTiO$_3$ (STO: bandgap 3.2~eV),
from the LaO interface layer to the TiO$_2$ layer. For a defect-free
interface, the concentration of these ``intrinsic'' sheet carriers is
$n_{\rm sheet}\approx 3.3 \times 10^{14} {\rm cm}^{-2}$. For samples
prepared under low oxygen pressure ($<10^{-4}$ mbar), the carrier
density is much larger than this intrinsic value, suggesting their
properties are determined by extrinsic carriers related to oxygen
vacancies formed in the STO substrate during the growth of the LAO
film.

\begin{figure}[t!]
\includegraphics[scale=0.40]{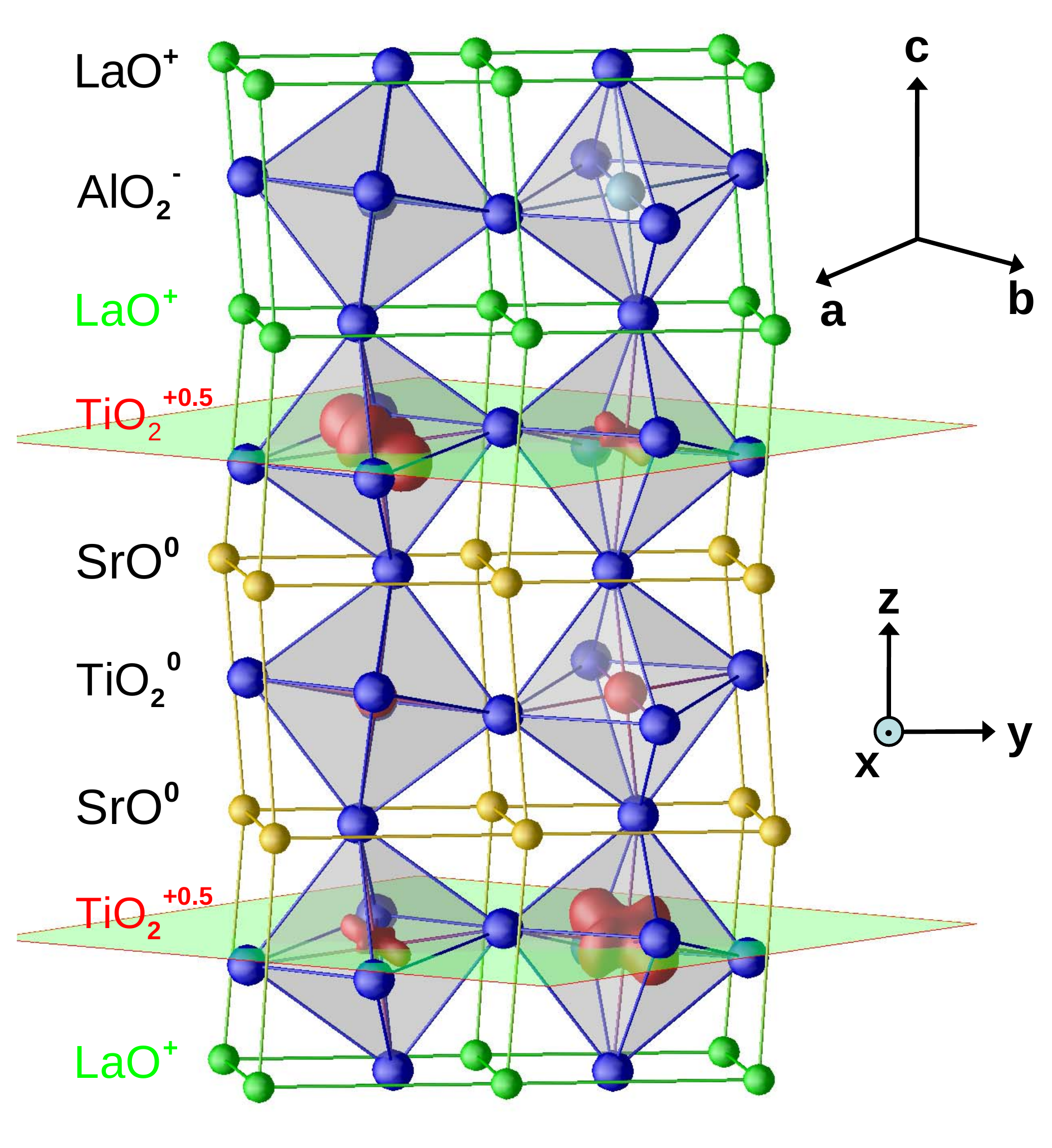}
\caption{Bulk LaTiO$_{3}$ based LaAlO$_3|$SrTiO$_3$ interface structure
and charge density iso-surface of the surplus electron for the charge
ordered ferromagnetic state. The orthorhombic translation vectors are
\textbf{a}, \textbf{b}, \textbf{c}.} \label{Fig1}
\end{figure}

Samples prepared under increasing oxygen pressure exhibit a large
increase of the sheet resistance
\cite{Ohtomo:nat04,Herranz:prl07,Brinkman:natm07}. A sheet resistance
which decreases on increasing the temperature from 10K to 70K, large
negative magnetoresistance and magnetic hysteresis at low temperatures
were found in Ref.~\cite{Brinkman:natm07} and these properties are
presumably characteristic of intrinsic interfaces. The partial filling
of the Ti $d$ states at the $n$-type interface means that locally the
electronic structure is intermediate between orthorhombic LaTiO$_3$
(Ti$^{3+}$) and cubic SrTiO$_3$ (Ti$^{4+}$). LaTiO$_3$ (LTO) is a
G-type antiferromagnetic (AFM) Mott insulator with a Neel temperature
of 146~K \cite{MacLean:jssc79,Cwik:prb03,Hemberger:prl03} and a band
gap of 0.2~eV \cite{Okimoto:prb95}. Compared to cubic STO, the
additional electron in the $t_{2g}$ conduction band state on the Ti ion
leads to a GdFeO$_3$-type crystal structure (space group \emph{Pbnm})
which can be derived from the ideal perovskite cubic structure by
tilting essentially ideal TiO$_6$ octahedra about the {\bf b}-axis
followed by a rotation about the {\bf c}-axis (see Fig.~\ref{Fig1}).
% The La ions are displaced mainly along the {\bf b} axis with a small
% component along the {\bf a} axis
% \cite{Mochizuki:prl03,Pavarini:prl04,Solovyev:prb96}.

In this paper we suggest that GdFeO$_3$-type distortions at LAO$|$STO
interfaces play an essential role in reducing the bandwidth of the
occupied interface Ti $d$ state, rendering it very sensitive to on-site
Coulomb correlations and stabilizing an antiferromagnetic ground state.
Surprisingly little attention has been paid to the possibility of such
distortions at the LAO$|$STO interface. Theoretical studies have been
largely concerned with explaining the metallic behavior of $n$-type and
the insulating character of $p$-type interfaces
\cite{Okamoto:nat04,Pentcheva:prb06,Park:prb06,Lee:prb07,Albina:prb07}
with most emphasis being placed on the charge transfer between the LaO
and TiO$_2$ interface layers.

Though LDA total energy calculations describe equilibrium crystal
structures with good accuracy and have been applied extensively to
non-metallic TM oxides \cite{Rabe:tap07}, we are not aware of
applications to the structures of open shell correlated solids. To
establish a predictive capability for the LAO$|$STO system, we carried
out a systematic study of the crystal structure of the 3$d^1$
perovskites CaVO$_3$, SrVO$_3$, YTiO$_3$ and LaTiO$_3$. We could show
that LDA calculations reproduce the well-documented
\cite{Pavarini:njp05} GdFeO$_3$-type distortions in this system very
well. To study various types of magnetic ordering, we took account of
the Coulomb interaction of the partially filled $d$-shell using a local
Coulomb interaction parameter $U_d$ \cite{Anisimov:jpcm97} which has
been shown to describe the ground state of YTiO$_3$ and LaTiO$_3$
qualitatively correctly \cite{Okatov:epl05}. The LDA$+U$ results are in
even closer agreement with experiment \cite{Zhong:prb08} .

\emph{Method:} We model the LAO$|$STO interface using a periodically
repeated ($m, n$) supercell containing $m$ layers of LAO and $n$ layers
of STO. Most of the results to be reported below were obtained with the
40 atom $c(2\times2)$ ($\frac{3}{2},\frac{5}{2}$) supercell depicted in
Fig.~\ref{Fig1} containing two $n$-type interfaces, and did not change
significantly when a ($\frac{5}{2},\frac{7}{2}$) supercell was used
instead. To study AFM ordering the above supercell was doubled in the
$xy$ plane leading to a $p(2\times2)$ unit cell containing 80 atoms.

The LDA and LDA$+U$ calculations were carried out using the projector
augmented wave (PAW) method \cite{Blochl:prb94b,Kresse:prb99} and a
plane wave basis as implemented in the Vienna Ab initio Simulation
Package (VASP) \cite{Kresse:prb93,Kresse:prb96}. A kinetic energy
cutoff of 500~eV was used and the Brillouin zone of the 40 atom
supercell was sampled with a 6$\times$6$\times$4 k-point grid in
combination with the tetrahedron method \cite{Blochl:prb94a}. Compared
to the LDA, the LDA$+U$ approach gives an improved description of $d$
electron localization \cite{Anisimov:jpcm97}. We use the rotationally
invariant LDA$+U$ method \cite{Dudarev:prb98} with, unless stated
otherwise, $U_d=5$ and $J_d=0.64$~eV for Ti $d$ states. The La $f$
states are forced to lie higher in energy by imposing a large
$U_f=11$~eV and $J_f=0.68$~eV \cite{Okamoto:prl06}.

\begin{figure}[!]
\includegraphics[scale=0.32, angle=0]{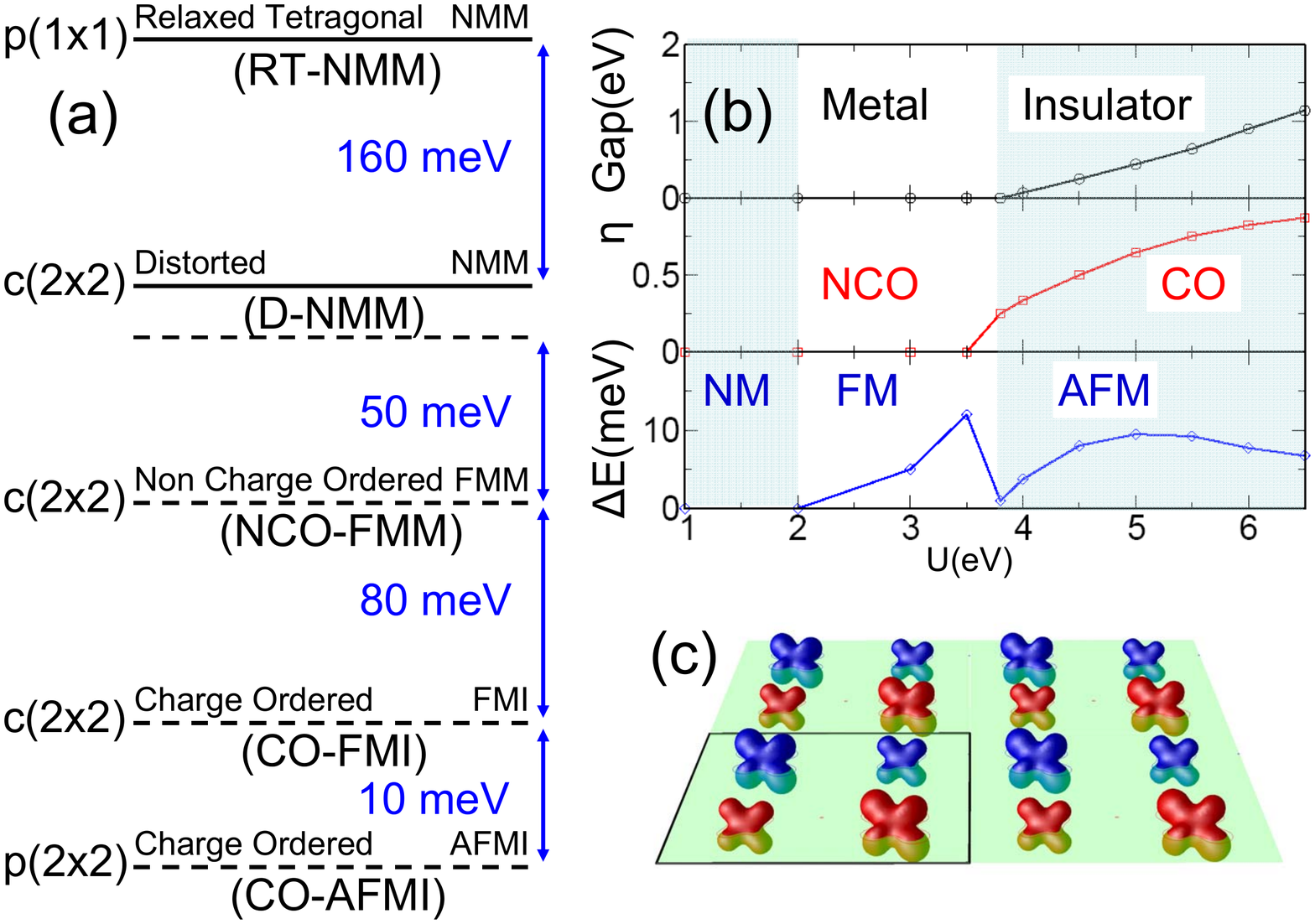}
\caption{ (a) Energy gained per surplus electron by relaxing symmetry
constraints. The LDA$+U$ energies (dashed lines) and energy differences
calculated here with $U_d=5$~eV depend on the value of $U$ used and so
are just indicative.
(b) Phase diagram as function of value of $U_d^{Ti}$. Top: energy gap.
Middle: disproportionation order parameter $\eta=|n_1 - n_2|/(n_1 +
n_2$) where $n_i$ is the occupation of the Ti $d$ states on ion $i$.
Bottom: $\Delta E = |E_{COFM}-E_{COAFM}|$.
(c) Spatial magnetic distribution of the COAFM states ($U_d=4$eV) at
the interface.} \label{Fig2}
\end{figure}

\emph{Results:} Because samples are grown on STO substrates, we fix the
in-plane lattice constant at the experimental value for STO. This
imposes a strain on LAO with a 3\% smaller bulk lattice constant. Our
starting point is a fully relaxed, tetragonal, non-magnetic metallic
(RT-NMM) $p(1\times1)$ configuration. We then remove symmetry
constraints to obtain, in order of increasing energy gain
(Fig.~\ref{Fig2}a):
a $c(2\times2)$ distorted non-magnetic metallic (D-NMM) structure;
a metastable spin-polarized, non charge-ordered (NCO) state, with a
single interface-Ti charge state, obtained by switching on $U_d^{Ti}$;
a charge-ordered ferromagnetic insulating (CO-FMI) state with an energy
gap of 0.44~eV in which the Ti sites are not equivalent
(Fig.~\ref{Fig1});
a $p(2\times2)$ charge-ordered anti-ferromagnetic insulating state
(CO-AFMI).

The charge transferred from LAO to the smaller bandgap STO is bound
close to the interface. As predicted by Okamoto and Millis
\cite{Okamoto:nat04} and confirmed for LTO$|$STO interfaces using
calculations similar to those presented here \cite{Okamoto:prl06}, the
effect of lattice relaxation is to smooth the potential discontinuity
and allow charge to leak into the bulklike STO. For our NCO-FMM state,
the occupied part of the bottom of the conduction band is almost 1eV
wide and we confirm the occurrence of charge leakage at the LAO$|$STO
interface.

\begin{figure}[!]
\includegraphics[scale=0.33, angle=0]{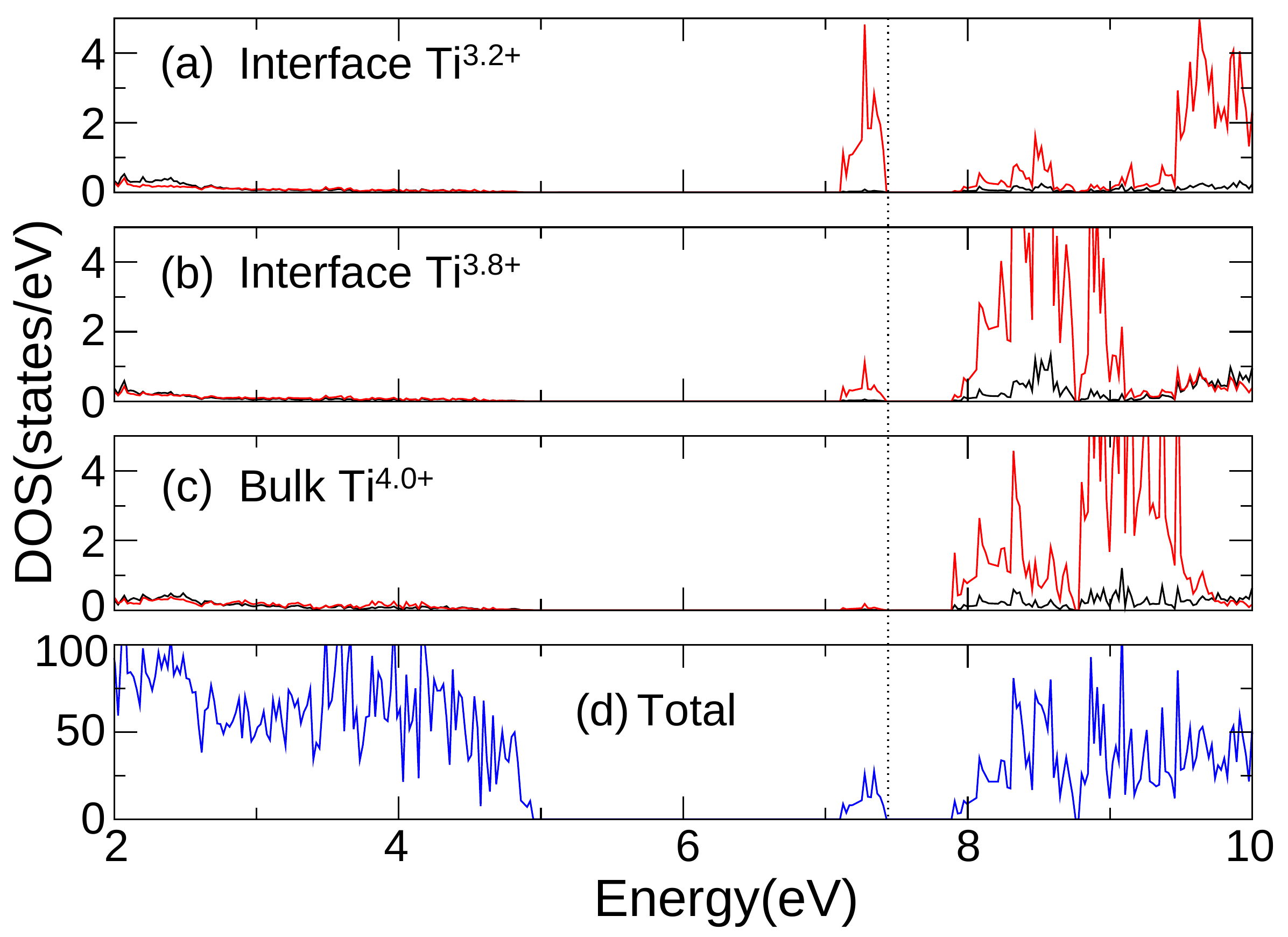}
\caption{Interface densities of states (DoS) for the lowest energy
charge ordered antiferromagnetic insulating (CO-AFMI) configuration.
(a) and (b) projected $3d$ DoS of (nominally) Ti$^{3+}$ and Ti$^{4+}$
interface Ti ions. The transferred electron has $t_{2g}$ (red)
character with the $e_{g}$ states (black) far above the Fermi energy;
(c) projected $3d$ DoS of bulk Ti$^{4+}$ ions; (d) total DoS.}
\label{Fig3}
\end{figure}

The NCO-FMM state is unstable with respect to charge disproportionation
into inequivalent Ti$^{3.2+}$ and Ti$^{3.8+}$ interface ions as a
result of the Coulomb repulsion and spin polarization on the Ti ions
favoring integer occupancy; the transferred electron is localized on
the nominally $3+$ interface Ti$^{3+}$ ions with little leakage into
bulk STO (Fig.~\ref{Fig3}). The atoms relax to accommodate the charge
ordering: small displacements of the oxygen atoms surrounding the
Ti$^{3+}$ ion lead to that TiO$_6$ octahedron expanding while the other
one contracts. The combination of this distortion with the potential
barrier and reduced symmetry at the interface leads to one of the Ti
$t_{2g}$ states splitting off completely from the bottom of the STO
conduction band and a gap opening up for quarter filling. The
transferred electron remains strongly localized at the interface with
essentially no leakage into the bulk layer. Because of the
GdFeO$_3$-type interface distortion, the occupied Ti $t_{2g}$ state has
$0.79|xy\rangle+0.33|yz\rangle+0.52|xz\rangle$ orbital character
(Fig.~\ref{Fig1}).

In a $p(2\times2)$ geometry, the kinetic energy can be further reduced
by flipping the spins of equivalent Ti ions in a checkerboard pattern.
The bandwidth is narrowed to 0.4~eV and the energy gap of the CO-AFMI
state increased by 20~meV compared to the CO-FMI state. The total
energy is lowered but the energy gain $\sim 10$~meV is small, close to
the limit of our accuracy.

The effect of varying $U_d^{Ti}$ between 2 and 7~eV, keeping $J$ fixed
at 0.64~eV, is shown in Fig.~\ref{Fig2}b. Between 3.5 and 4~eV, charge
disproportionation takes place, AF ordering becomes more favourable and
a metal insulator transition occurs. Increasing $U_d$ leads to a larger
CO gap and with increasing charge disproportionation the stripe charge
and AF ordered pattern shown in Fig.~\ref{Fig2}c evolves into a
checkerboard pattern. The energy difference between CO-FMI and CO-AFMI
states reaches a maximum at $U_d=5$eV, the effect of increasing $U$
being to suppress the exchange coupling between neighbouring spins. The
most important result of this study is that the TiO$_6$ rotation
reduces the bandwidth of the split-off band so that very modest values
of $U_d$ result in FM and AFM ordering and a metal insulator (MI)
transition close to the FM-AFM phase boundary.

\emph{Structure:} Because the LAO lattice constant is 3\% smaller than
that of STO, LAO might be expected to shrink in the $z$ direction when
it is forced to match in-plane to STO. However, at the interface, we
find that the La and Sr planes are separated by 4.003{\AA} which is
substantially larger than $(3.905 + 3.789)/2$, the arithmetic average
of the LAO and STO lattice constants, and somewhat larger than the
lattice constant of a fictitious cubic LTO phase with the same volume
as the experimentally observed orthorhombic phase, but consistent with
recent experiments \cite{Vonk:prb07}. By comparing results for
($\frac{3}{2},\frac{5}{2}$) and ($\frac{5}{2},\frac{7}{2}$) multilayers
for the NCO-FMM and CO-FMI states, we find that the GdFeO$_3$ rotation
of the TiO$_6$ octahedra is confined to the interface. The central STO
(LAO) layers in the ($\frac{5}{2},\frac{7}{2}$) multilayers are
essentially cubic (tetragonal). Compared to a bulk orthorhombic
LaTiO$_3$ with an appropriately scaled lattice parameter, the shift of
the La and Sr ions is reduced, presumably as a result of the
constraints imposed by $\frac{5}{2}$ or $\frac{7}{2}$ layers of cubic
STO and the availability of only half an extra electron in the Ti $d$
states.

\begin{figure}[!]
\includegraphics[scale=0.35, angle=0]{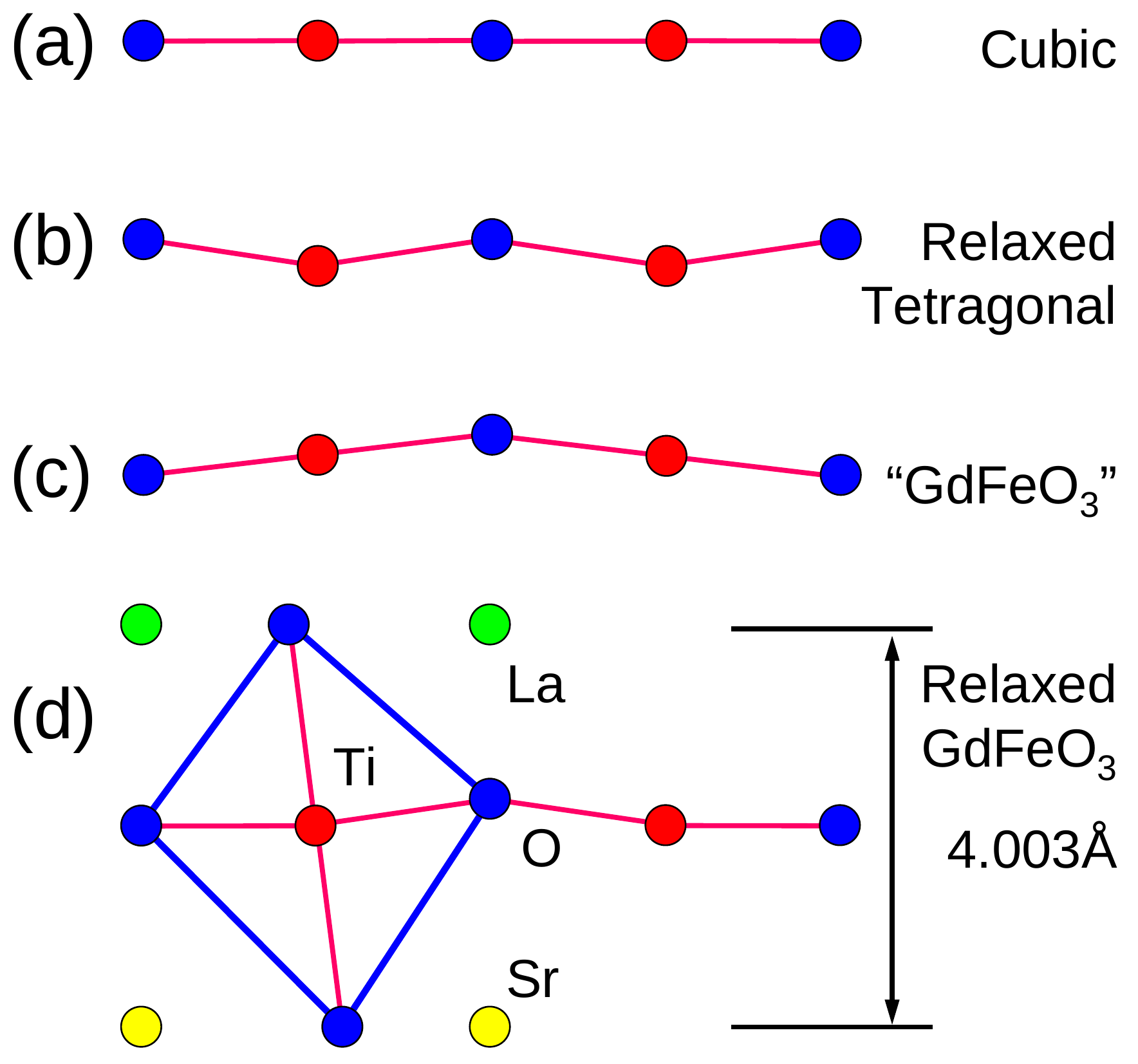}
\caption{Sketch of the vertical displacement patterns of the Ti and O
interface atoms for (a) an ideal cubic structure (b) a relaxed
$p(1\times1)$ structure (c) an ideal $c(2\times2)$ ``GdFeO$_3$''
structure and (d) the fully relaxed $c(2\times2)$ ``buckling''
structure. } \label{structure2}
\end{figure}

Previous LDA and LDA$+U$ calculations found ferroelectric-like
displacements of negatively charged O and positively charged Ti ions
when the geometry was optimized for $p(1\times1)$ structures
\cite{Park:prb06,Okamoto:prl06} resulting in a zigzag pattern of
O-Ti-O-Ti-O Ti displacements perpendicular to the interface; see
Fig.~\ref{structure2}b. Combining this with the GdFeO$_3$ rotation of
the TiO$_6$ octahedra (Fig.~\ref{structure2}c) leads to the buckling
pattern shown in Fig.~\ref{structure2}d in which one oxygen is
displaced by $\sim0.15$\AA\ out of an otherwise essentially linear
chain.

\emph{Discussion:} Let us assume that we are just on the insulating
side of the MI transition in Fig.~\ref{Fig2}b i.e., $U_d^{Ti}\sim
4$~eV. Our calculations predict that the strong correlation of Ti $d$
states will lead to electrons transferred from the LaO to the TiO$_2$
layer being trapped on the interface Ti ions in an AFMI state. The
localization of these intrinsic carriers would explain the high sheet
resistance ($>10^{4}\Omega/\Box$) and low carrier concentration ($\sim
10^{13}{\rm cm}^{-2}$) of samples prepared at high oxygen pressure
\cite{Ohtomo:nat04,Brinkman:natm07,Siemons:prl07,Herranz:prl07}. The
existence of a band gap would explain the decrease of sheet resistance
with increasing temperature \cite{Brinkman:natm07}. Application of an
external magnetic field would force the AFMI system into a
ferromagnetic insulating state with a smaller band gap, leading to a
reduced resistance as observed.

This cannot be the whole picture, however. For example, a simple
(doped) semiconductor model would predict carrier freeze-out at low
temperatures and this is not observed. The role of oxygen vacancies as
well as that of the cation mixing at the interface needs to be
clarified. If we focus on the ideal of an intrinsic, abrupt interface,
perhaps the most pressing issue to be resolved experimentally relates
to the appropriate value of $U_d^{Ti}$ which is a free parameter in our
otherwise parameter-free study. A spectroscopic study identifying the
existence and size of the predicted CO gap would be invaluable for
making further progress.

\emph{Summary and Conclusions:} Using LDA and LDA$+U$ calculations, we
have shown that the half electron (per interface Ti ion) transferred
from the LaO layer on one side of an LAO$|$STO interface to the TiO$_2$
layer on the other side favors a rotation of TiO$_6$ octahedra just as
it does in bulk LTO. As in bulk LTO, the distortion is crucial to the
formation of a charge-ordered AFMI ground state and other charge,
magnetic and orbital properties, and results in a characteristic
buckling of the Ti-O-Ti bonding at the interface. Ionic relaxation
plays a crucial role in determining the localization of the extra
electron on the interface Ti ions and leakage into the bulk layer.
However, charge ordering suppresses this leakage even when relaxation
is included and, in the CO states, the electrons are strongly localized
at the interface. Our calculations suggest an explanation for recent
experimental results in terms of the proximity of an AFMI ground state
to a FMM excited state (or a FMI state with reduced band gap). With
only half an electron trapped in an otherwise empty Ti $d$ orbital, the
LAO$|$STO interface is an attractive object for quarter-filled Hubbard
model studies.

\emph{Acknowledgments:} This work is supported by ``NanoNed'', a
nanotechnology programme of the Dutch Ministry of Economic Affairs. Use
of supercomputer facilities was sponsored by the ``Stichting Nationale
Computer Faciliteiten'' (NCF) financially supported by the
``Nederlandse Organisatie voor Wetenschappelijk Onderzoek'' (NWO). The
authors thank P.X. Xu, Q.F. Zhang, D. Vanpoucke, S. Kumar, A. Brinkman
and G. Rijnders for useful discussions.

\end{document}